\documentclass[11pt,a4paper]{article}

\usepackage[margin=2.5cm]{geometry}
\usepackage{color}
\usepackage{amsmath,amssymb,amsthm,amsopn,bm,mathrsfs}
\usepackage{setspace}
\usepackage{graphbox}
\usepackage[en-GB]{datetime2}
\DTMlangsetup[en-GB]{ord=omit}
\usepackage{algorithm}
\usepackage{algorithmicx}
\usepackage[noend]{algpseudocode}
\usepackage{natbib}
\usepackage{booktabs}
\usepackage{makecell} 
\usepackage[hidelinks]{hyperref} 

\DeclareMathOperator{\E}{\boldsymbol{\mathbb{E}}}
\DeclareMathOperator{\Prob}{\boldsymbol{\mathbb{P}}}
\DeclareMathOperator{\Err}{Err}

\newcommand{\bg}{{\bm g}}
\newcommand{\bx}{{\bm x}}
\newcommand{\bX}{{\bf X}}
\newcommand{\diff}{\mathsf{d}}
\newcommand{\pkg}[1]{{\tt #1}}
\newcommand{\code}[1]{{\tt #1}}
\newcommand{\indi}[1]{\boldsymbol{1} \lbrace #1 \rbrace}

\newsavebox\CBox
\def\textBF#1{\sbox\CBox{#1}\resizebox{\wd\CBox}{\ht\CBox}{\textbf{#1}}}

\title{Interpretable contour level selection for heat maps for gridded data}

\author{Tarn Duong\footnote{Paris, France F-75000. Email:
{\tt tarn.duong@gmail.com}. ORCID: \href{https://orcid.org/0000-0002-1198-3482}{0000-0002-1198-3482}.}}

\begin{document}

\maketitle

\begin{abstract}
\noindent Gridded data formats, where the observed multivariate data are aggregated into grid cells, ensure confidentiality and reduce storage requirements, with the trade-off that access to the underlying point data is lost. Heat maps are a highly pertinent visualisation for gridded data, and heat maps with a small number of well-selected contour levels offer improved interpretability over continuous contour levels. There are many possible contour level choices. Amongst them, density contour levels are highly suitable in many cases. Current methods for computing density contour levels requires access to the observed point data, so they are not applicable to gridded data. To remedy this, we introduce an approximation of density contour levels for gridded data. We then compare our proposed method to existing contour level selection methods, and conclude that our proposal provides improved interpretability for synthetic and experimental gridded data.
\end{abstract}

\maketitle

\section{Introduction}

Gridded data are an effective format for the storage and dissemination of confidential and high-volume data. Confidential data are aggregated within a grid cell to ensure their privacy, whereas high volume data (aka Big data) are aggregated into grid cells to reduce their storage requirements.
A heat map is a common visualisation method for bivariate gridded data where the value of the variable of interest is encoded by a colour scale. 
A continuous colour scale provides the most complete visualisation of the data \citep{tobler1973}, though the large number of colours may impede interpretability \citep{jenks1963,dobson1973}. 
The usual approach to increasing interpretability is to reduce the number of distinct colours by assigning the same colour to an interval of values.
The key question is to select an optimal set of contour levels and a colour scale to describe meaningfully the structure of the data surface \citep{klemela2009}. 
For simplicity, we treat the colour scale selection here as a secondary question, since we follow the recommendations in \cite{zeileis2009}, and focus our attention on the contour level selection. 

There are many possibilities for the contour level selection: for example, the `natural' contour levels \citep{jenks1967} are widely employed. Despite their widespread acceptance, the spatial relevance of their associated contour regions has not been established rigorously within a statistical framework. In contrast, density contour levels rigorously define spatially relevant regions \citep{bowman1993jasa,hyndman1996jasa}. A $\tau$-density contour level is associated with the smallest region that contains $\tau$ probability mass of the data sample.  
Whilst the probabilistic interpretation of the density contour regions is crucial for interpreting density estimates, we claim that density contour levels can be extended so that they are also useful for the visualisation of gridded data which are not density estimates. Though we continue to call them `density contour levels' as this is already well-established.  
The most popular method for calculating these density contour levels requires access to original point data which does not pertain to our case \citep{hyndman1996jasa}. Thus we propose an approximation of density contour levels which is suitable for gridded data.  

\begin{figure}[!ht]
\centering
\includegraphics[width=0.7\textwidth]{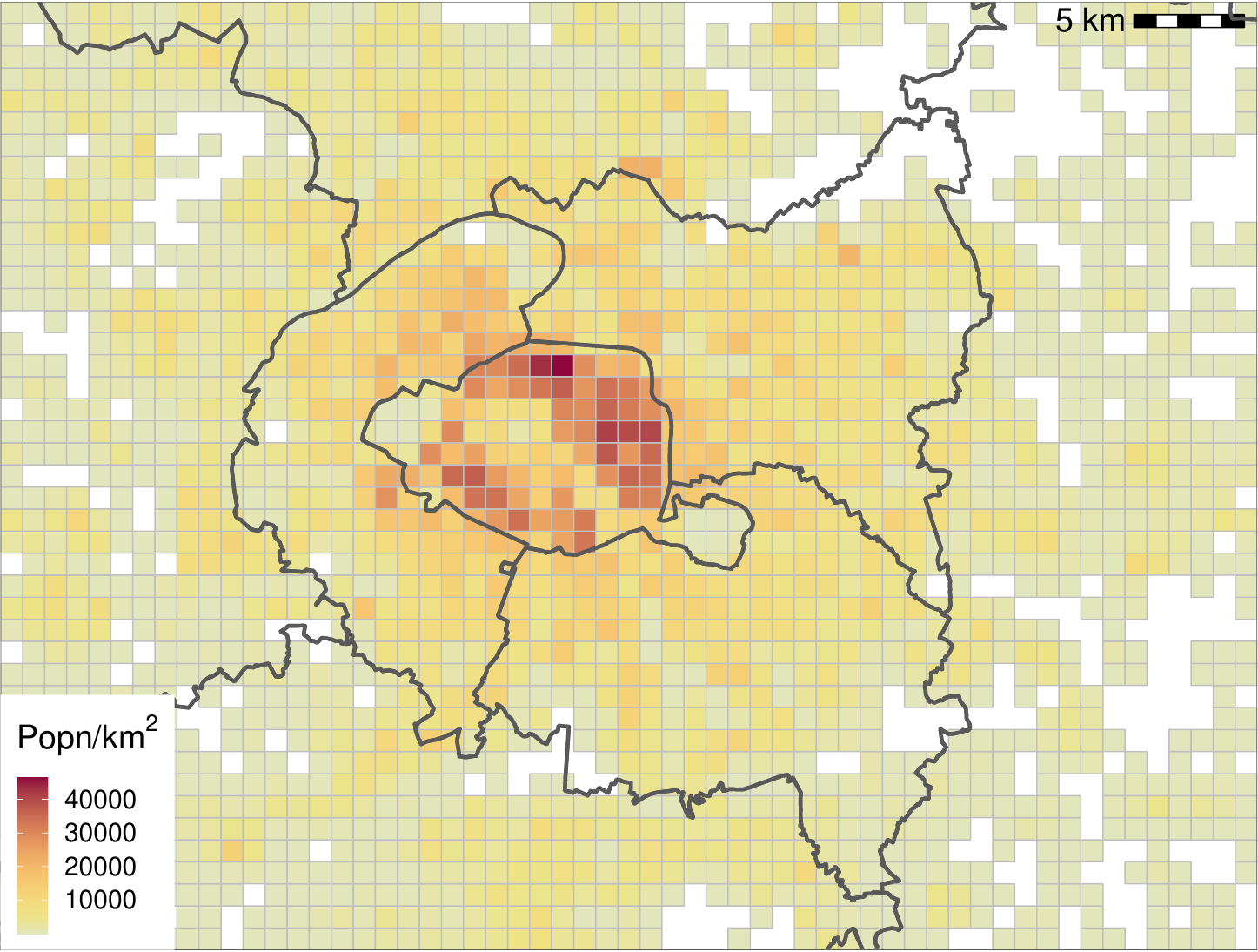} 
\caption{Heat map for geospatial gridded data. Population of the Paris city region on 1~km $\times$ 1~km grid, with continuous sequential `Heat' colour scale.} 
\label{fig:gridded1}
\end{figure}

Our first motivating data set is a 1~km $\times$ 1~km gridded data set of socioeconomic indicators from the 2019 household census of about 30 million households in France \citep{insee2019}. The socioeconomic indicators involve income, poverty and housing etc. Due to their sensitive nature, this grid resolution is a trade-off between the dissemination of detailed socio-economic indicators for informed policy decision making, and the guarantee of privacy for individual households. 
We focus on the population density in the ROI (region of interest) $[1.97^\circ\mathrm{E}, 2.79^\circ\mathrm{E}] \times [48.63^\circ \mathrm{N}, 49.06^\circ \mathrm{N}]$ which consists of 2\,078 grid cells and includes the Paris city centre and its inner ring of suburbs, as displayed in Figure~\ref{fig:gridded1}, in the EPSG 3035 projected coordinate system. The continuous colour scale, ranging from 40\,000 persons per km$^2$ (red) to 20\,000 persons per km$^2$ (orange) to 10\,000 persons per km$^2$ (yellow),  provides a visually appealing heat map of the population distribution. We are able to identify qualitatively the densely populated regions (hotspots), though it is challenging to compute quantitatively these hotspots.

\begin{figure}[!ht]
\centering
\includegraphics[width=\textwidth]{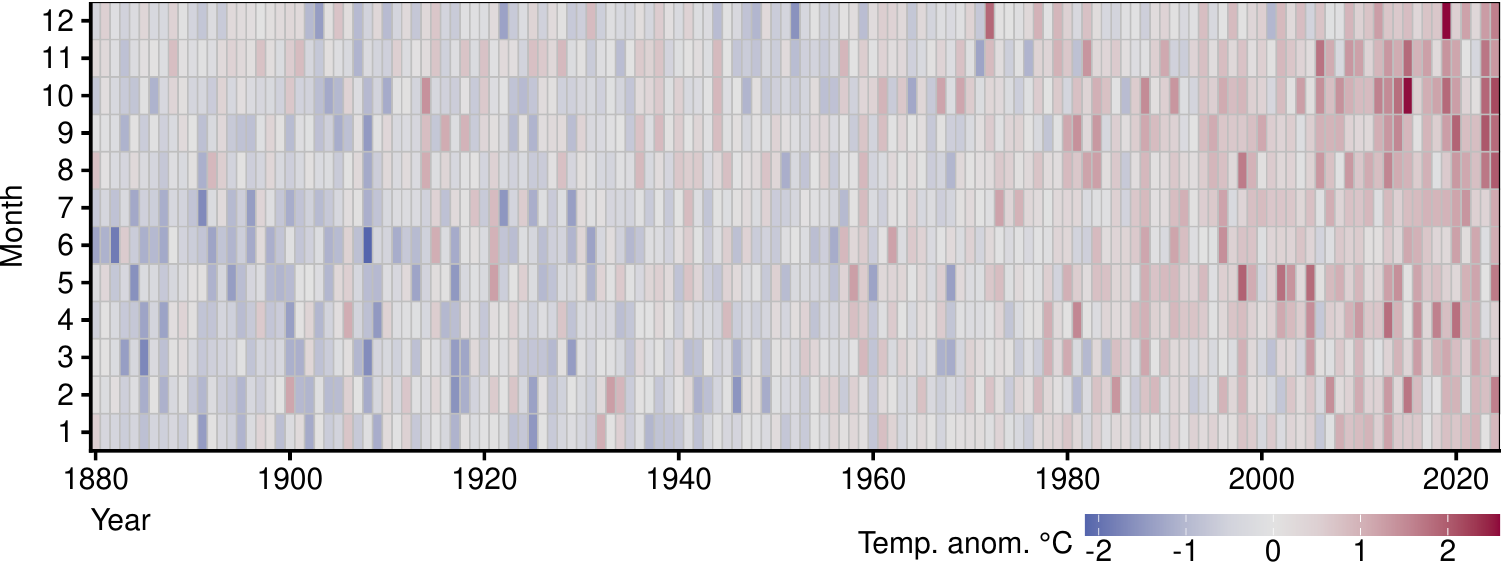} 
\caption{Heat map for non-geospatial gridded data. Year-monthly surface temperature anomaly time series in Western Australia, with continuous diverging `Red-Blue' colour scale.}
\label{fig:gridded2}
\end{figure}

Our second motivating data set is non-geospatial gridded data. It consists of the year, the month and the mean monthly  GISTEMP temperature anomaly, for all years from 1880 to 2024. It is a reconstructed data set, drawing from many high-volume point data sources, and the aggregation into grid cells is carried out to ensure comparable statistical reliability for all grid cells, so that it can serve as a global reference data source \citep{vose2021,lenssen2024,gistemp2025}. We focus on the temperature anomalies in the ROI $[111.28^\circ\mathrm{E}, 129.43^\circ\mathrm{E}] \times [35.79^\circ\mathrm{S}, 12.68^\circ\mathrm{S}]$ which covers the state of Western Australia. It comprises 1740 grid cells for each year-month combination from 1880-01 to 2024-12. The heat map plots the year on the horizontal axis and the month on the vertical axis. The long term trends in Figure~\ref{fig:gridded2} is from cool anomalies (blue) in the first half of the year in the 1880--1900s to hot anomalies (red) in the second half of the year since the 2000s, though the visual impression of these trends is not overly strong due to the continuous colour scale.  

This manuscript is organised as follows. In Section~\ref{sec:dens-cont}, we define rigorously density contour levels, and introduce our proposed approximation of density contour levels for gridded data. In Section~\ref{sec:data-analysis}, we verify the efficacy of this approximation for synthetic and experimental gridded data, as well as the increased interpretability of density contour levels over existing contour level estimation methods. We end with some concluding remarks.

\section{Density contours} \label{sec:dens-cont}

We begin by examining the special case for a probability density function. A $\tau$-density contour $R_\tau$ is defined as the smallest region that contains the upper $\tau$ probability of the density function  $f$ 
\begin{equation} \label{eq:dens-cont}
R_\tau  = \{ \bx : f(\bx) \geq f_\tau \}  \ \mathrm{and} \ \Prob(\bX \in R_\tau) = \tau 
\end{equation}
where the random variable $\bX$ is drawn from $f$ \citep{bowman1993jasa,hyndman1996jasa}. 
This definition of $R_\tau$ in Equation~\eqref{eq:dens-cont} involves the implicitly defined $\tau$-density contour level $f_\tau$. 
\cite{hyndman1996jasa} asserts an alternative explicit definition of $f_\tau$ as the
$(1-\tau)$-quantile of $Y= f(\bX)$, i.e. 
\begin{equation} \label{eq:dens-cont-level} 
f_\tau =  F_Y^{-1} (1-\tau)
\end{equation}
where $F_Y$ is the cumulative distribution function of $Y$. 
This quantile is well-defined as $Y$ is a univariate random variable. Then,  
$\Prob(\bX \in R_\tau) = \int_{\mathbb{R}^d} f(\bx) \indi{f(\bx) \geq f_\tau} \, \diff\bx 
= \E [\indi{f(\bX) \geq f_\tau}]
= \Prob (f(\bX) \geq f_\tau)
= \Prob(Y \geq f_\tau) = 1- F_Y(f_\tau)  
= 1 - F_Y( F_Y^{-1}(1-\tau)) = \tau$, which verifies the probability associated with $R_\tau$, i.e. the $f_\tau$ in Equations~\eqref{eq:dens-cont} and \eqref{eq:dens-cont-level} are the same. The proof that $R_\tau$ has the minimal hypervolume amongst all sets of probability mass  $\tau$ is more involved and the interested reader is urged to consult \citet{hyndman1996jasa}. We call $R_\tau$ a `density contour region', following \citet{polonik1995,chacon2018}, whilst \citet{hyndman1996jasa} prefers `highest density region'. 

A set of $m$ density contour levels $0 < \tau_1 < \dots < \tau_m < 1$, and their corresponding $m$ density contour regions $R_{\tau_1}, \dots, R_{\tau_m}$ facilitate an intuitive interpretation of density visualisations. 
Common choices include the quartile contours $\tau = 0.25, 0.5, 0.75$ or the odd decile contours $\tau = 0.1, 0.3, 0.5, 0.7, 0.9$, or the decile contours $\tau = 0.1, \dots, 0.9$, depending on the intricacy of structure of the density $f$. The upper limit of a suitable number of contour levels is most likely about eight since this approaches the limit of human abilities for visual differentiation \citep{jenks1963,delicado2015}. 

\begin{figure}[!ht]
\centering
\includegraphics[width=0.9\textwidth]{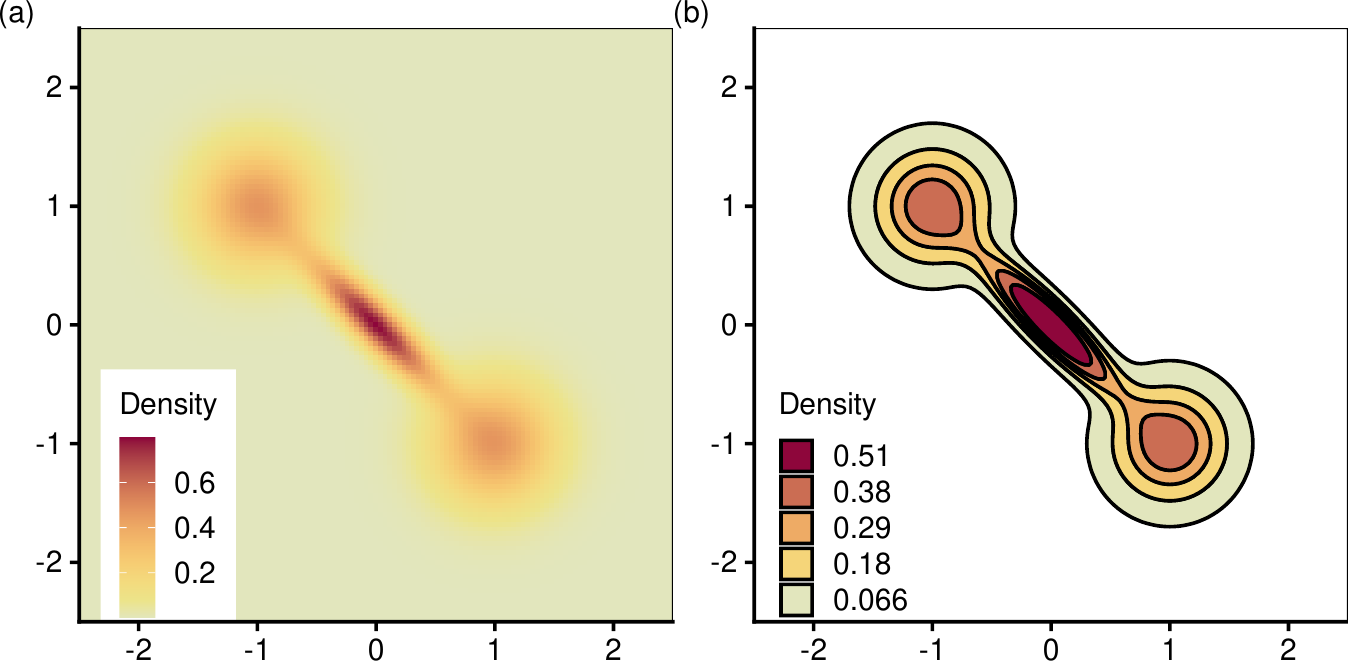}
\caption{Gaussian mixture density $4/11 N((-1,1), 1/8[1, 0; 0, 1]) + 3/11 N((0,0), 1/8[1, 9/10; 9/10, 1]) + 4/11 N((1,-1), 1/8[1, 0; 0, 1])$. 
(a) Heat map  with continuous `Heat' colour scale, red (high) to orange (mid) to yellow (low). (b) Heat map with decile contour levels.}
\label{fig:dens-cont}
\end{figure}

In Figure~\ref{fig:dens-cont}, we illustrate the additional interpretability of density contour levels over a continuous colour scale for a multimodal Gaussian mixture density $3/11 N((-1,1), 1/8[1, 0; 0, 1]) + 4/11 N((0,0), 1/8[1, 9/10; 9/10, 1]) + 3/11 N((1,-1), 1/8[1, 0; 0, 1])$. 
In Figure~\ref{fig:dens-cont}(a) is the heat map of the density function with a continuous colour scale, from ranging red (high density) to orange (mid) to yellow (low). We are able to ascertain some qualitative characteristics, such as the multimodality and the relative heights/locations of the modes, they are not evident. 
In Figure~\ref{fig:dens-cont}(b), the solid black curves are the decile contour regions $f_{0.9}=0.066 , f_{0.7} =0.18, f_{0.5} = 0.29, f_{0.3} = 0.38, f_{0.1}=0.51$. With this set of contour levels, the structure of the density function is more clearly ascertained.  

The density contour level $f_\tau$ in Equation~\eqref{eq:dens-cont-level} remains difficult to compute since it requires a closed form of both the density $f$ and the cumulative distribution of $f(\bX)$. The mostly commonly deployed method is a quantile-based approximation from a random sample $\bX_1, \dots, \bX_n$ drawn from the common density $f$ \citep{hyndman1996jasa}.  From this sample data, we construct an estimator $\hat{f}$ of the target density $f$. Within a parametric framework, maximum likelihood estimators are popular, and within a non-parametric framework, histograms and kernel density estimators are widely used. The estimation methodology does not need to be specified here as the following apply for any consistent density estimator. The $\tau$-density contour level $f_\tau$ is consistently estimated by plug-in estimators of Equations~(\ref{eq:dens-cont}--\ref{eq:dens-cont-level}), i.e., $\tilde{f}_\tau$ is the sample $(1-\tau)$-quantile of $\hat{f}(\bX_1), \dots, \hat{f}(\bX_n)$. This approach is outlined in Algorithm~\ref{alg:dens-contq}. The inputs are the random sample of size $n$ and the probability $\tau$. The output is the approximate $\tau$-density contour level $\tilde{f}_\tau$. In Step~1, we draw a random data sample $\bX_1, \dots, \bX_n$ of size $n$ from the common density $f$. In Step~2, we compute the density estimate $\hat{f}$ from the data sample. In Step~3, we compute the transformed data sample by evaluating the density estimate at the data sample values. In Step~4, the density contour level is approximated as the $(1-\tau)$-quantile of these transformed data values.  

\begin{algorithm}[!ht]
\caption{Quantile-based approximation of density contour level}
\label{alg:dens-contq}
\begin{algorithmic}[1]
\Statex {\bf Input:} random sample $\bX_1, \dots, \bX_n$, $n$ sample size, $\tau$ probability
\Statex {\bf Output:} $\hat{f}_\tau$ approximate density contour level 
\State Compute density estimate $\hat{f}$ from $\bX_1, \dots, \bX_n$
\State Evaluate density estimate at data sample values $\hat{f}(\bX_1), \dots, \hat{f}(\bX_n)$
\State Approximate density contour level $\tilde{f}_\tau := (1-\tau)$-quantile of $\hat{f}(\bX_1), \dots, \hat{f}(\bX_n)$
\end{algorithmic}
\end{algorithm}

The calculations in Algorithm~\ref{alg:dens-contq} are intuitive, efficient, and valid for any data dimension $d$ and any consistent density estimator $\hat{f}$, which has contributed to its wide acceptance. However we cannot employ this algorithm for gridded data since it requires direct access to the original point data sample. 
So we search for alternatives to Steps~2--3 which avoid explicit computations with the point data $\bX_1, \dots, \bX_n$.   

For our proposed alternative in Algorithm~\ref{alg:dens-cont}, 
the inputs are the density estimate $\hat{f}$, a set of $M$ grid points $G = \{\bg_1, \dots, \bg_M\}$,  and the probability $\tau$. 
In Step~1, we compute the hypervolume $\delta$ of a grid cell from $G$. 
In Steps~2--4, we compute the $M$ probability elements for the grid $G$: the $j$th probability element $p_j$ is obtained by multiplying the hypervolume $\delta$ and the density estimate value $\hat{f}(\bg_j)$ at the grid point $\bg_j$. 
In Steps~5--6, we compute the order statistics of the $M$ probability elements, and their partial sums. In Step~7, we search for the smallest cell index $j^*$ such that the partial sum is greater than or equal to $\tau$, i.e. $P_{j^*} = p_{(j^*)} + \dots + p_{(M)}$ exceeds $\tau$ by the least amount. By construction, the union  $\hat{R}_\tau = \bg_{(j^*)} \cup \cdots \cup \bg_{(M)}$ is the smallest region, comprised of the grid cells, with probability mass $\geq \tau$, i.e. $\hat{R}_\tau$ is an approximation of $R_\tau$. In Step~8, the density contour level is approximated by $\hat{f}(\bg_{j^*})$, which corresponds to $\tilde{f}_\tau$ from Algorithm~\ref{alg:dens-contq}.

\begin{algorithm}[!ht]
\caption{Grid-based approximation of density contour level}
\label{alg:dens-cont}
\begin{algorithmic}[1]
\Statex {\bf Input:} $\hat{f}$ density estimate, $G$ estimation grid, $\tau$ probability 
\Statex {\bf Output:} $\tilde{f}_\tau$ approximate density contour level
\State Compute hypervolume of grid cell $\delta$   
\For{$j := 1$ to $M$}
\State Evaluate density estimate at grid point $\hat{f}(\bg_j)$
\State Compute probability element $p_j := \delta \hat{f}(\bg_j)$ 
\EndFor
\State Compute order statistics $p_{(1)} \leq \cdots \leq p_{(M)}$ of probability elements 
\State Compute partial sums $P_j := \sum_{\ell=j}^M p_{(\ell)}$ of order statistics
\State  Find the smallest index $j^* \in \{1, \dots, M\}$ such that $P_{j^*}  \geq \tau$
\State Approximate density contour level $\hat{f}_\tau := \hat{f} (\bg_{j^*})$
\end{algorithmic}
\end{algorithm}
  
Due to the invariance of density contour levels, Algorithm~\ref{alg:dens-cont} can be adapted to any non-negative function $g$ with a finite integral over the grid $G$. With $\hat{g}$, a grid-based estimate of $g$, we calculate $S$ which is the sum of all probability elements of $\hat{g}$ on the grid $G$. 
If we normalise the grid-based values as $\hat{f} = \hat{g}/S$, then $\hat{f}$ is a suitable density estimate input in  Algorithm~\ref{alg:dens-cont}. Since the order statistics of $p_1/S, \dots, p_M/S$ remain the same as those for $p_1, \dots, p_M$, then the approximate contour level for $g$ is $\hat{g}_\tau = S\hat{f}_\tau$. This is the situation for the Paris population data. 

Furthermore, for a function $g$ whose range includes negative and positive values, then we partition $g = g^+ - g^-$, where $g^-(\bx) = -\indi{g(\bx) \leq 0} g(\bx)$ and $g^+(\bx) = \indi{g(\bx) > 0} g(\bx)$, where $\indi{\cdot}$ is the indicator function. Since $g^-, g^+$ are non-negative functions, then we can apply Algorithm~\ref{alg:dens-cont} to each of them, resulting in contour levels $\hat{g}^-_\tau, \hat{g}^+_\tau$ for each of the negative and positive values of $g$, yielding
the contour regions $\{ \bx : \hat{g}(\bx) \leq \tilde{g}^-_\tau\}$  and $\{ \bx : \hat{g}(\bx) \geq \hat{g}^+_\tau\}$ respectively. Whilst these contour regions do not have a probabilistic interpretation,  $\hat{g}^-_\tau, \hat{g}^+_\tau$ remain suitable contour level choices for visualising $\hat{g}$. Thus Algorithm~\ref{alg:dens-cont} can be adapted to compute contour levels for any grid-based function, and is not restricted to density functions.  This is the situation for the temperature anomaly data. 

We conclude with a brief description of several other contour level selection methods. The naive quantile contour levels compute the $\tau$-quantile of all values in the grid $\hat{f}(\bg_1), \dots,  \hat{f}(\bg_M)$. Suppose that we select $m$ naive quantile contour levels, then for comparison, we can compute the $m$ equal interval length contour levels, namely $\hat{f}_{\min} + j/m \cdot (\hat{f}_{\max} - \hat{f}_{\min}), j= 1, \dots, m$, or the $m$ `natural' or Jenks contour levels \citep{jenks1967}, which is a $k$-means clustering with $m$ clusters of $\hat{f}(\bg_1), \dots,  \hat{f}(\bg_M)$. These contour level selection methods are easy to understand and to implement.  
In the following section, we compare the empirical performance of these contour level selection methods to density contour levels for synthetic and experimental data.

\section{Data analysis} \label{sec:data-analysis}

To demonstrate the applicability of density contour levels for gridded data, we deploy the visualisation capabilities of the \pkg{R} statistical analysis environment. Due to the simplicity of Algorithm~\ref{alg:dens-cont}, it can be implemented in any visualisation software.

\subsection{Synthetic data}

\begin{figure}[!htp]
\centering
\includegraphics[height=0.9\textheight]{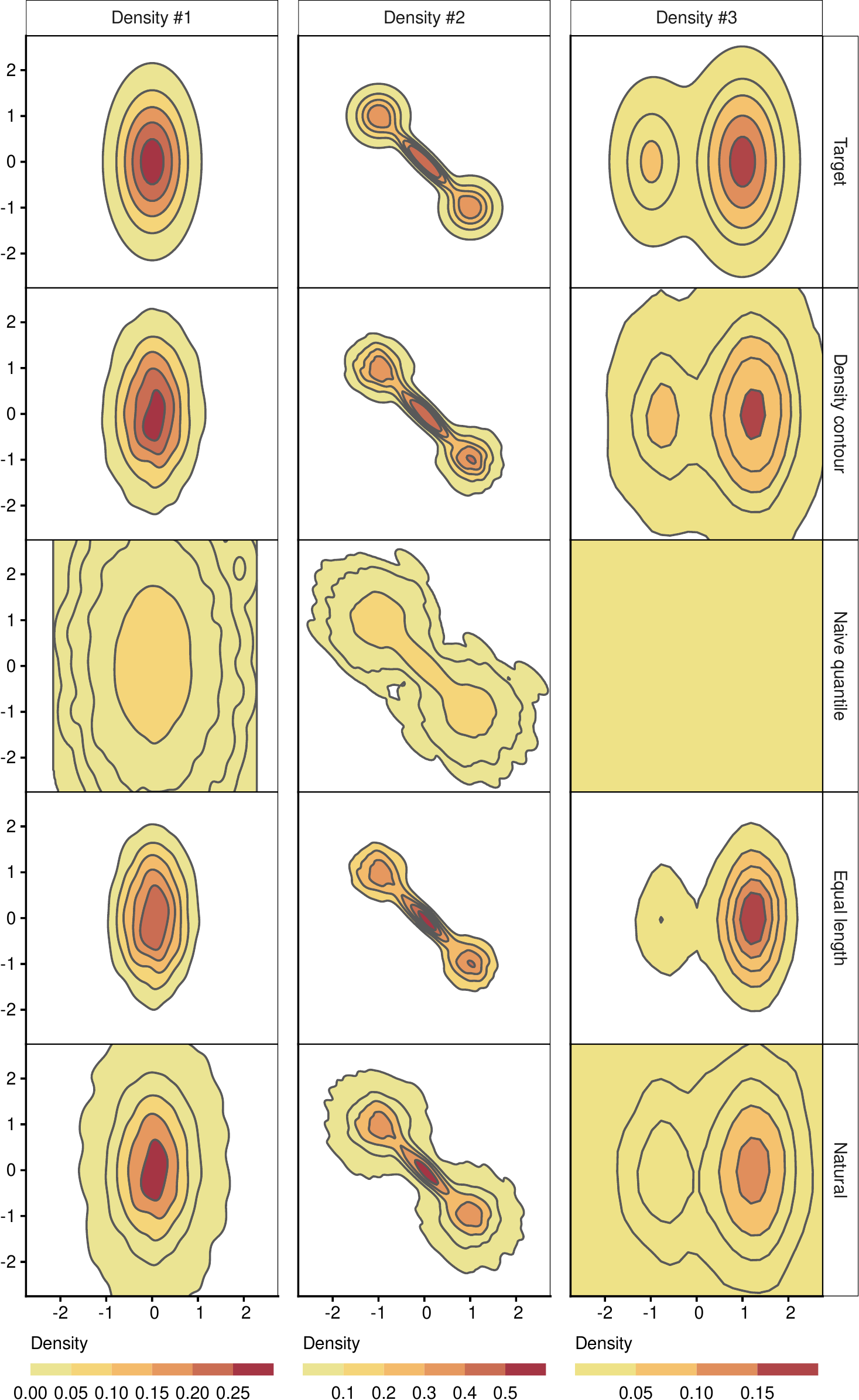}
\caption{Comparison of contour level selection methods for gridded density estimates from mixture densities, with  discretised sequential `Heat' colour scale, with $n=10\,0000$ sample. (First row) Target proxy contour. (Second row) Density contour.  (Third row) Naive quantile. (Fourth row) Equal length. (Fifth row) Natural (Jenks).} 
\label{fig:mixt}
\end{figure}

We focus on two bivariate Gaussian mixture densities and a $t$-mixture density: Density~\#1 is a base case $N((-1,0)$, [1/4, 0; 0, 1]); Density~\#2 is trimodal with a central mode oriented obliquely from the coordinates axes $4/11 N((-1,1), 1/8[1, 0; 0, 1]) + 3/11 N((0,0), 1/8[1, 9/10; 9/10, 1]) + 4/11 N((1,-1), 1/8[1, 0; 0, 1])$; Density~\#3 is bimodal with asymmetric modes and heavy tails $1/4 t((-1,0), [1/4, 0; 0, 1], 10) + 3/4 t((1,0), [1/4, 0; 0, 1], 10)$, where each $t$ component has 10 d.f. 
For each density, we draw a Monte Carlo $N=10^6$ random sample $\bX_1, \dots, \bX_N$, and we transform this random sample using the target mixture density $f$ to obtain $f(\bX_1), \dots, f(\bX_N)$. The quantile-based approximation $f^*_\tau$ is the upper $\tau$-quantile of  $f(\bX_1), \dots, f(\bX_N)$, from which we can establish $R^*_\tau=\{\bx : f(\bx) \geq f^*_\tau\}$. We use $f^*_\tau, R^*_\tau$ as the proxies for $f_\tau, R_\tau$. These $R^*_\tau, \tau=0.1, 0.3, 0.5, 0.7, 0.9$ are displayed in the first row in Figure~\ref{fig:mixt}.

We then draw another $n=10\,000$ random sample from the target mixture density, apply Algorithm~\ref{alg:dens-contq} with a kernel density estimate $\hat{f}$, on an $M=151 \times 151$ grid, to yield quantile-based estimates $\tilde{f}_\tau$ and $\tilde{R}_\tau$. The kernel density estimate is computed using a plug-in bandwidth \citep{duong2003}, though any sensible density estimate can be used here. To mimic the situation where we only have access to the gridded density estimate and we do not have access to the data sample, then we input only the grid of the previous kernel density estimate values $\hat{f}(\bg_1), \dots, \hat{f}(\bg_M)$ into Algorithm~\ref{alg:dens-cont} to yield the grid-based approximations $\hat{f}_\tau$ and $\hat{R}_\tau$. The approximation computations are carried out by the \code{as.kde} function in the \pkg{ks} package \citep{ks} or by the \code{tidy\_as\_kde}/\code{st\_as\_kde} functions in the \code{eks} package \citep{eks}.
These $\hat{R}_\tau$ are the second row in Figure~\ref{fig:mixt}. 
We observe that the contour region estimates $\hat{R}_\tau$ generally remain close to $R^*_\tau$. 
In Figure~\ref{fig:mixt}, the third row is the naive quantiles, the fourth row the equal length intervals, and the fifth the natural contour levels. Overall, we observe that the density contours are the closest to the target proxy contours, followed by the natural and equal length contours, and then the naive quantile contours. 

If we focus on Density \#3, the proxy contour levels are $f^*_{0.9} = 0.014, f^*_{0.7}=0.041, f^*_{0.5}=0.071, f^*_{0.3}=0.13, f^*_{0.1} = 0.20$. The density contour levels are $\hat{f}_{0.9} = 0.006, \hat{f}_{0.7}= 0.028, \hat{f}_{0.5} =0.051, \hat{f}_{0.3}=0.088, \hat{f}_{0.1} =0.17$, and they describe the bimodality structure and account for the heavy tails well.     
The naive quantile contour levels are $0, 0, 1.87{\rm e-}19, 7.08{\rm e-}19, 2.26{\rm e-}5$, as the vast majority of the gridded estimate values are close to zero; the contour plot with these contour regions does not reveal any structure. The equal length contour levels are $0.033, 0.066, 0.10, 0.13, 0.17$, which describes both modes less well, especially the smaller mode. The natural contour levels $5.58{\rm e-}5, 0.013, 0.036, 0.072, 0.14$, reveal some bimodality but suffer in comparison to the density contour and equal length levels. Natural contour levels are the results of a $k$-means clustering, and it well-known that this is not robust to large departures from Gaussianity and is sub-optimal in the presence of the heavy tails of a $t$-mixture. 

Since Figure~\ref{fig:mixt} gives us good reason to believe that the proxy contours provide a good description of the structure of the underlying density, then we employ them for a quantitative comparison. That is, the target are the proxy decile contour regions $R^*_\tau, \tau=0.1, 0.3, 0.5, 0.7, 0.9$.   
To explore the performance of the contour level estimation methods, we consider two grid sizes $M=51 \times 51$ and $M=151 \times 151$, and two sample sizes $n=1\,000$ and $n=10\,000$. For each combination of these parameter settings, we draw 100 replicates from the target mixture density. For each replicate, we compute the (i) density contour regions from Algorithm~\ref{alg:dens-cont} with the same $\tau$ values, (ii) naive quantile contour regions with the same $\tau$ values, (iii) 5 equal length contour regions, and (iv) 5 natural contour regions. The equal length and natural contour levels estimates are not based on quantiles, so we estimate 5 of them each to align with the number of $\tau$ levels. Since these estimated contour regions are set estimates, the usual measure of their error to the target contour region $R^*_\tau$ is the probability of the symmetric difference between $\hat{R}_\tau$ and $R_\tau^*$. That is, 
$\Err(\hat{R}_\tau) = \Prob (\bX \in \hat{R}_\tau \Delta R_\tau^*)$, where $A \Delta B  = A \cup B \backslash (A \cap B)$ is the symmetric difference between sets $A, B$. These probabilities are approximated as Riemann sums. 
 
These errors for the mixture densities are displayed in Table~\ref{tab:mixt-err1} ($n=1000$) and Table~\ref{tab:mixt-err2} ($n=10~000$). Within each sample size, we present the grid sizes $51 \times 51$ and $151 \times 151$. Within each density, each row in the table corresponds to a contour region estimation method: density contour, naive quantile, equal length, natural. The columns are the target proxy contour regions $R_\tau^*, \tau=0.9, 0.7, 0.5, 0.3, 0.1$. Each table entry is the mean $\pm$ standard deviation of the $\Err$ of the symmetric difference of estimated and target contour regions.

Overall, the results for $n=1000$ and $n=10~000$ are broadly similar, with the larger sample size resulting in errors with lower mean and with smaller spread, all other factors being equal. The density contour estimates have the smallest errors in the majority of cases. In the remaining minority of cases, they are usually second by a small margin. On the other hand, the naive quantile estimates are almost always the worst, and in some cases, by a large margin. The equal length and natural contours can be effective in some cases, with a slight advantage for the natural contour regions.    
 
\begin{table}[ht]
\centering\setlength{\tabcolsep}{4pt}
\begin{tabular}{@{}clrrrrrr@{}}
\toprule
\thead{Density} & \thead{Estimator} & \thead{$R_{0.9}^*$} & \thead{$R_{0.7}^*$} & \thead{$R_{0.5}^*$} & \thead{$R_{0.3}^*$} & \thead{$R_{0.1}^*$} \\ 
\midrule
\multicolumn{7}{@{}l}{$\mathit{n=1000, M=51 \times 51}$} \\ 
\#1 & Density contour & \textBF{0.058$\pm$0.022} & \textBF{0.083$\pm$0.026} & 0.084$\pm$0.023 & 0.072$\pm$0.018 & 0.040$\pm$0.010 \\ 
    & Naive quantile  & 0.602$\pm$0.029 & 0.417$\pm$0.031 & 0.224$\pm$0.030 & 0.070$\pm$0.019 & 0.100$\pm$0.010 \\ 
    & Equal length    & 0.088$\pm$0.028 & 0.099$\pm$0.027 & 0.085$\pm$0.022 & \textBF{0.067$\pm$0.018} & 0.049$\pm$0.010 \\ 
    & Natural         & 0.088$\pm$0.021 & 0.120$\pm$0.018 & \textBF{0.083$\pm$0.021} & 0.068$\pm$0.019 & \textBF{ 0.036$\pm$0.009} \\[0.5ex] 
\#2 & Density contour & \textBF{0.103$\pm$0.029} & 0.238$\pm$0.056 & 0.269$\pm$0.074 & 0.224$\pm$0.063 & 0.135$\pm$0.036 \\ 
    & Naive quantile  & 0.735$\pm$0.021 & 0.572$\pm$0.028 & 0.432$\pm$0.037 & 0.262$\pm$0.048 & 0.130$\pm$0.033 \\ 
    & Equal length    & 0.082$\pm$0.025 &\textBF{0.216$\pm$0.041} & \textBF{0.259$\pm$0.067} & \textBF{0.217$\pm$0.063} & 0.123$\pm$0.035 \\ 
    & Natural & 0.112$\pm$0.041 & 0.223$\pm$0.047 & 0.265$\pm$0.066 & 0.223$\pm$0.064 & \textBF{0.129$\pm$0.038} \\[0.5ex]
\#3 & Density contour & \textBF{0.059$\pm$0.024} & \textBF{0.081$\pm$0.029} & 0.093$\pm$0.026 & \textBF{0.095$\pm$0.024} & \textBF{0.041$\pm$0.010} \\ 
    & Naive quantile  & 0.635$\pm$0.040 & 0.458$\pm$0.042 & 0.266$\pm$0.042 & 0.105$\pm$0.031 & 0.095$\pm$0.014 \\ 
    & Equal length    & 0.064$\pm$0.024 & 0.079$\pm$0.027 & \textBF{0.088$\pm$0.026} & 0.114$\pm$0.025 & 0.133$\pm$0.023 \\ 
    & Natural         & 0.070$\pm$0.023 & 0.100$\pm$0.026 & 0.132$\pm$0.031 & 0.112$\pm$0.029 & 0.059$\pm$0.019 \\ 
\midrule
\multicolumn{7}{@{}l}{$\mathit{n=1000, M=151 \times 151}$}\\ 
\#1 & Density contour & \textBF{0.042$\pm$0.016} & \textBF{0.052$\pm$0.014} & \textBF{0.050$\pm$0.011} & 0.045$\pm$0.009 & 0.029$\pm$0.007 \\  
    & Naive quantile  & 0.600$\pm$0.028 & 0.412$\pm$0.028 & 0.216$\pm$0.029 & 0.044$\pm$0.015 & 0.094$\pm$0.010 \\ 
    & Equal length    & 0.075$\pm$0.029 & 0.082$\pm$0.034 & 0.055$\pm$0.018 & 0.041$\pm$0.009 & 0.045$\pm$0.012 \\ 
    & Natural         & 0.072$\pm$0.018 & 0.120$\pm$0.018 & 0.059$\pm$0.015 & \textBF{0.039$\pm$0.008} & \textBF{0.024$\pm$0.005} \\[0.5ex] 
\#2 & Density contour & 0.038$\pm$0.016 & \textBF{0.123$\pm$0.025} & \textBF{0.117$\pm$0.018} & 0.097$\pm$0.013 & 0.064$\pm$0.009 \\ 
    & Naive quantile  & 0.706$\pm$0.020 & 0.517$\pm$0.020 & 0.357$\pm$0.021 & 0.164$\pm$0.020 & 0.080$\pm$0.026 \\ 
    & Equal length    & \textBF{0.035$\pm$0.013} & 0.136$\pm$0.025 & 0.136$\pm$0.036 & 0.094$\pm$0.021 & 0.060$\pm$0.013 \\ 
    & Natural         & 0.041$\pm$0.024 & 0.142$\pm$0.028 & 0.141$\pm$0.033 & \textBF{0.086$\pm$0.016} & \textBF{0.049$\pm$0.008} \\[0.5ex] 
\#3 & Density contour & \textBF{0.035$\pm$0.013} & \textBF{0.044$\pm$0.012} & 0.062$\pm$0.012 & \textBF{0.063$\pm$0.013} & \textBF{0.029$\pm$0.006} \\ 
    & Naive quantile  & 0.638$\pm$0.047 & 0.454$\pm$0.048 & 0.259$\pm$0.047 & 0.082$\pm$0.029 & 0.087$\pm$0.015 \\ 
    & Equal length    & 0.040$\pm$0.015 & 0.053$\pm$0.016 & \textBF{0.058$\pm$0.015} & 0.088$\pm$0.022 & 0.141$\pm$0.025 \\ 
    & Natural         & 0.043$\pm$0.014 & 0.068$\pm$0.026 & 0.107$\pm$0.025 & 0.112$\pm$0.029 & 0.065$\pm$0.020 \\ 
\bottomrule
\end{tabular}
\caption{Errors of estimates of contour regions, for mixture densities with sample size $n=1000$, and grid sizes $M= 51 \times 51, 151 \times 151$. Within each density, each row is contour region estimation method: density contour, naive quantile, equal length, natural. The columns are proxy contour regions $R_\tau^*, \tau=0.9, 0.7, 0.5, 0.3, 0.1$. Each table entry is the mean $\pm$ SD of error of symmetric difference of estimated and target contour regions, bold entry has smallest mean error.}
\label{tab:mixt-err1}
\end{table}

\begin{table}[ht]
\centering\setlength{\tabcolsep}{4pt}
\begin{tabular}{@{}clrrrrrr@{}}
\toprule
\thead{Density} & \thead{Estimator} & \thead{$R_{0.9}^*$} & \thead{$R_{0.7}^*$} & \thead{$R_{0.5}^*$} & \thead{$R_{0.3}^*$} & \thead{$R_{0.1}^*$} \\ 
\midrule
\multicolumn{7}{@{}l}{$\mathit{n=10\,000, M=51 \times 51}$} \\ 
\#1 & Density contour & \textBF{0.042$\pm$0.017} & \textBF{0.058$\pm$0.025} & \textBF{0.058$\pm$0.025} & 0.049$\pm$0.020 & \textBF{0.026$\pm$0.009} \\ 
   & Naive quantile   & 0.657$\pm$0.025 & 0.463$\pm$0.025 & 0.262$\pm$0.024 & 0.076$\pm$0.019 & 0.098$\pm$0.006 \\ 
   & Equal length     & 0.079$\pm$0.020 & 0.071$\pm$0.023 & 0.060$\pm$0.024 & 0.049$\pm$0.019 & 0.053$\pm$0.005 \\ 
   & Natural          & 0.079$\pm$0.014 & 0.119$\pm$0.015 & 0.064$\pm$0.020 & \textBF{0.048$\pm$0.020} & 0.027$\pm$0.008 \\[0.5ex]  
\#2 & Density contour & 0.112$\pm$0.015 & \textBF{0.235$\pm$0.035} & 0.286$\pm$0.042 & 0.232$\pm$0.034 & 0.122$\pm$0.020 \\ 
    & Naive quantile  & 0.751$\pm$0.015 & 0.577$\pm$0.019 & 0.433$\pm$0.022 & 0.257$\pm$0.024 & 0.129$\pm$0.022 \\ 
    & Equal length    & \textBF{0.085$\pm$0.015} & 0.244$\pm$0.033 & \textBF{0.262$\pm$0.034} & \textBF{0.229$\pm$0.034} & 0.122$\pm$0.016 \\ 
    & Natural         & 0.103$\pm$0.019 & 0.264$\pm$0.034 & 0.267$\pm$0.037 & 0.231$\pm$0.034 & \textBF{0.120$\pm$0.020} \\[0.5ex] 
\#3 & Density contour & \textBF{0.065$\pm$0.025} & \textBF{0.087$\pm$0.033} & 0.094$\pm$0.032 & \textBF{0.088$\pm$0.029} & \textBF{0.037$\pm$0.012} \\ 
   & Naive quantile   & 0.745$\pm$0.032 & 0.561$\pm$0.031 & 0.367$\pm$0.030 & 0.170$\pm$0.026 & 0.062$\pm$0.027 \\ 
   & Equal length     & 0.070$\pm$0.026 & 0.091$\pm$0.026 & \textBF{0.092$\pm$0.030} & 0.100$\pm$0.026 & 0.150$\pm$0.011 \\ 
   & Natural          & 0.075$\pm$0.026 & 0.105$\pm$0.027 & 0.127$\pm$0.027 & 0.119$\pm$0.028 & 0.061$\pm$0.014 \\ 
\midrule
\multicolumn{7}{@{}l}{$\mathit{n=10\,0000, M=151 \times 151}$} \\
\#1 & Density contour & \textBF{0.027$\pm$0.011} & \textBF{0.032$\pm$0.010} & \textBF{0.029$\pm$0.009} & 0.025$\pm$0.006 & \textBF{0.015$\pm$0.003} \\ 
   & Naive quantile  & 0.644$\pm$0.024 & 0.447$\pm$0.024 & 0.246$\pm$0.024 & 0.053$\pm$0.020 & 0.093$\pm$0.006 \\ 
   & Equal length    & 0.074$\pm$0.023 & 0.053$\pm$0.022 & 0.031$\pm$0.010 & 0.028$\pm$0.008 & 0.056$\pm$0.006 \\ 
   & Natural         & 0.067$\pm$0.012 & 0.130$\pm$0.011 & 0.056$\pm$0.011 & \textBF{0.024$\pm$0.006} & 0.025$\pm$0.004 \\[0.5ex] 
\#2 & Density contour & \textBF{0.029$\pm$0.015} & \textBF{0.068$\pm$0.022} & \textBF{0.073$\pm$0.023} & \textBF{0.057$\pm$0.016} & \textBF{0.030$\pm$0.006} \\ 
    & Naive quantile  & 0.709$\pm$0.018 & 0.501$\pm$0.019 & 0.334$\pm$0.019 & 0.134$\pm$0.020 & 0.085$\pm$0.017 \\ 
    & Equal length    & 0.029$\pm$0.014 & 0.076$\pm$0.024 & 0.128$\pm$0.022 & 0.092$\pm$0.019 & 0.075$\pm$0.009 \\ 
    & Natural         & 0.036$\pm$0.010 & 0.076$\pm$0.028 & 0.155$\pm$0.030 & 0.061$\pm$0.017 & 0.033$\pm$0.007 \\[0.5ex] 
\#3 & Density contour & \textBF{0.026$\pm$0.010} & \textBF{0.031$\pm$0.010} & \textBF{0.039$\pm$0.008} &\textBF{0.037$\pm$0.009} & \textBF{0.016$\pm$0.003} \\ 
    & Naive quantile  & 0.736$\pm$0.033 & 0.544$\pm$0.033 & 0.347$\pm$0.033 & 0.147$\pm$0.033 & 0.064$\pm$0.026 \\ 
    & Equal length    & 0.030$\pm$0.012 & 0.049$\pm$0.011 & 0.043$\pm$0.009 & 0.051$\pm$0.012 & 0.165$\pm$0.013 \\ 
    & Natural         & 0.032$\pm$0.010 & 0.045$\pm$0.009 & 0.077$\pm$0.013 & 0.125$\pm$0.012 & 0.080$\pm$0.013 \\ 
\bottomrule
\end{tabular}
\caption{Errors of estimates of contour regions, for mixture densities with sample size $n=10~000$, and grid sizes $M= 51 \times 51, 151 \times 151$. Within each density, each row is contour region estimation method: density contour, naive quantile, equal length, natural. The columns are proxy contour regions $R_\tau^*, \tau=0.9, 0.7, 0.5, 0.3, 0.1$. Each table entry is the mean $\pm$ SD of error of symmetric difference of estimated and target contour regions, bold entry has smallest mean error.}
\label{tab:mixt-err2}
\end{table}

Lastly, the finer $151 \times 151$ grids lead to lower errors than the coarser $51 \times 51$ grids, though the magnitude of the error reduction for an increased grid size depends on the data sample. For example, the largest decreases are for Density \#2 with its intricate multimodality, since a finer grid is able to more closely follow this complicated structure. Whilst a larger grid size would be better than a coarser grid in general, we recall that the grid size of gridded data is usually decided by the data provider. So the grid size for data analysis is not a tuning parameter that can be changed easily like in this simulation study.

\subsection{Experimental data}

For our analysis of experimental data, we return to the gridded data from Figures~\ref{fig:gridded1}-- \ref{fig:gridded2}. 
For the Paris population data, there are 2078 1~km $\times$ 1~km square polygons in the Lambert Azimuthal Equal Area Europe (EPSG 3035) projected coordinates system.
Since we rely on software which requires density estimates on uniform rectilinear grids aligned to the coordinate axes, so we interpolate them to a uniform rectangular grid which is aligned to the coordinate axes. For the population data, since it is mostly a uniform rectangular grid, we are only required to create the missing grid cells with population zero: the result is a grid of $M=57 \times 43 = 2451$ grid cells (1~km $\times$ 1~km). 
 
\begin{figure}[!ht]
\centering
\includegraphics[width=0.7\textwidth]{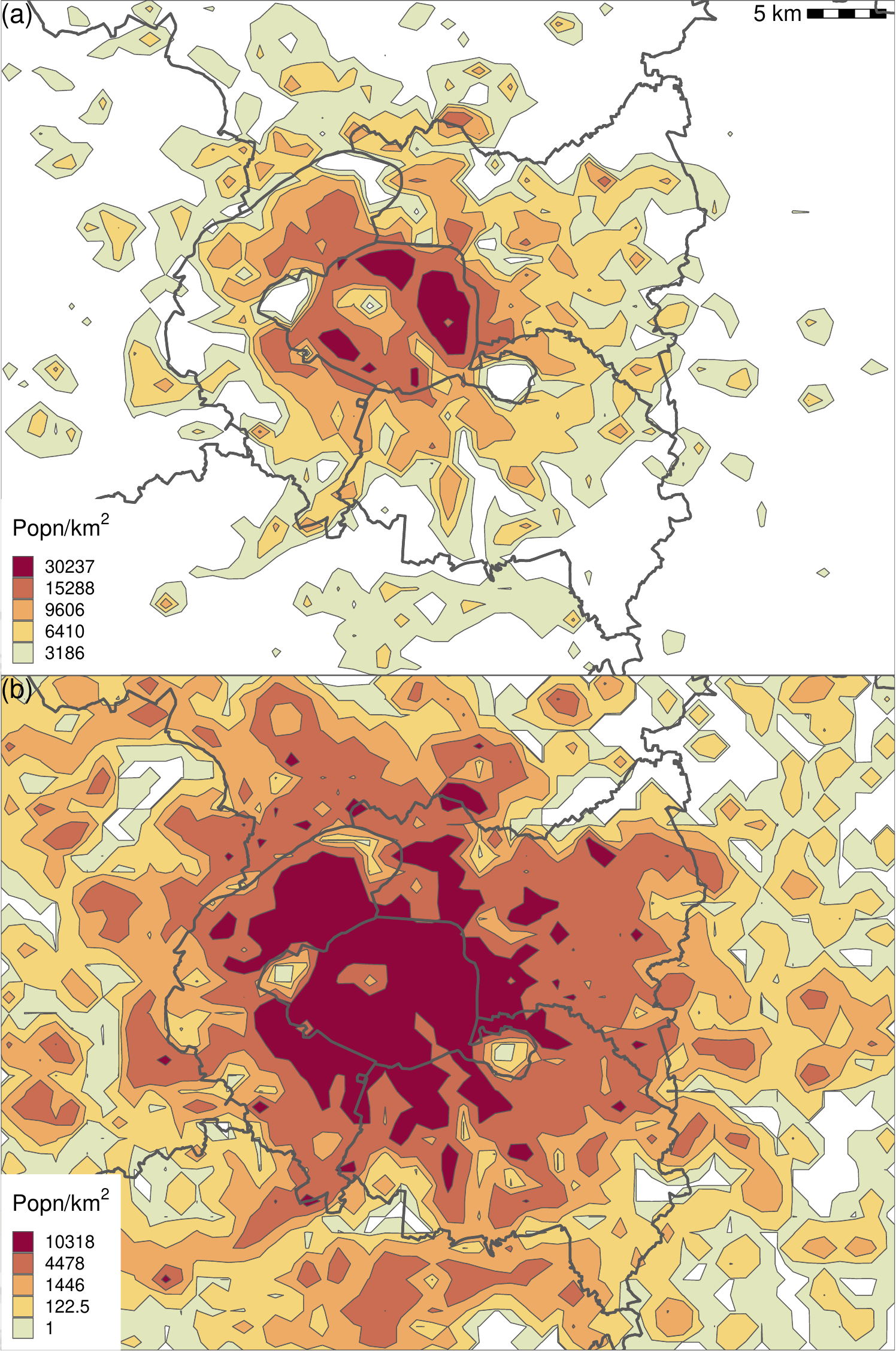}
\caption{Contour regions for population density for Paris capital region on 1~km $\times$ 1~km grid, with  discretised sequential `Heat' colour scale. (a) Density contour levels. (b) Naive quantile contour levels.}
\label{fig:idf-popn2}
\end{figure}

In Figure~\ref{fig:idf-popn2}(a) are the contour regions of the Paris population interpolated gridded data corresponding to the density contour levels: $\hat{f}_{0.1}=30\,237$ (red), $\hat{f}_{0.3}=15\,288$ (dark orange), $\hat{f}_{0.5}=9\,606$ (mid orange), $\hat{f}_{0.7}=6\,410$ (light orange), $\hat{f}_{0.9}=3\,186$ (yellow). In comparison to the continuous colour scale, these contour regions provide a more interpretable visualisation of the population distribution. 
The population hotspots can be more quantitatively defined as, say, the 10\% density contour region $\hat{R}_{0.1}$. That is, we can consider any area with population density of at least 30\,237 per km\textsuperscript{2} to be densely populated with respect to the rest of the ROI. The area of $\hat{R}_{0.1}$ is 22.29 km\textsuperscript{2}, which is 0.91\% of the total ROI area of 2451 km\textsuperscript{2}. Whereas the population in $\hat{R}_{0.1}$ is 957\,473.5 which is 10.09\% of the total ROI population of 9\,490\,878. 
These contour regions quantify the highly unequal population spatial distribution in the ROI. In contrast, the naive quantile contour levels are 10\,318 (red), 4\,478 (dark orange), 1\,446 (mid orange), 122 (light orange), 1 (yellow) in Figure~\ref{fig:idf-popn2}(b). Since zero or low population grid cells are by far the most numerous, even though their sum does not rival the population sum of the less numerous densely populated grid cells. So these contour regions give the erroneous impression that the population distribution is more evenly spread throughout the ROI.  

\begin{figure}[!ht]
\centering
\includegraphics[width=0.7\textwidth]{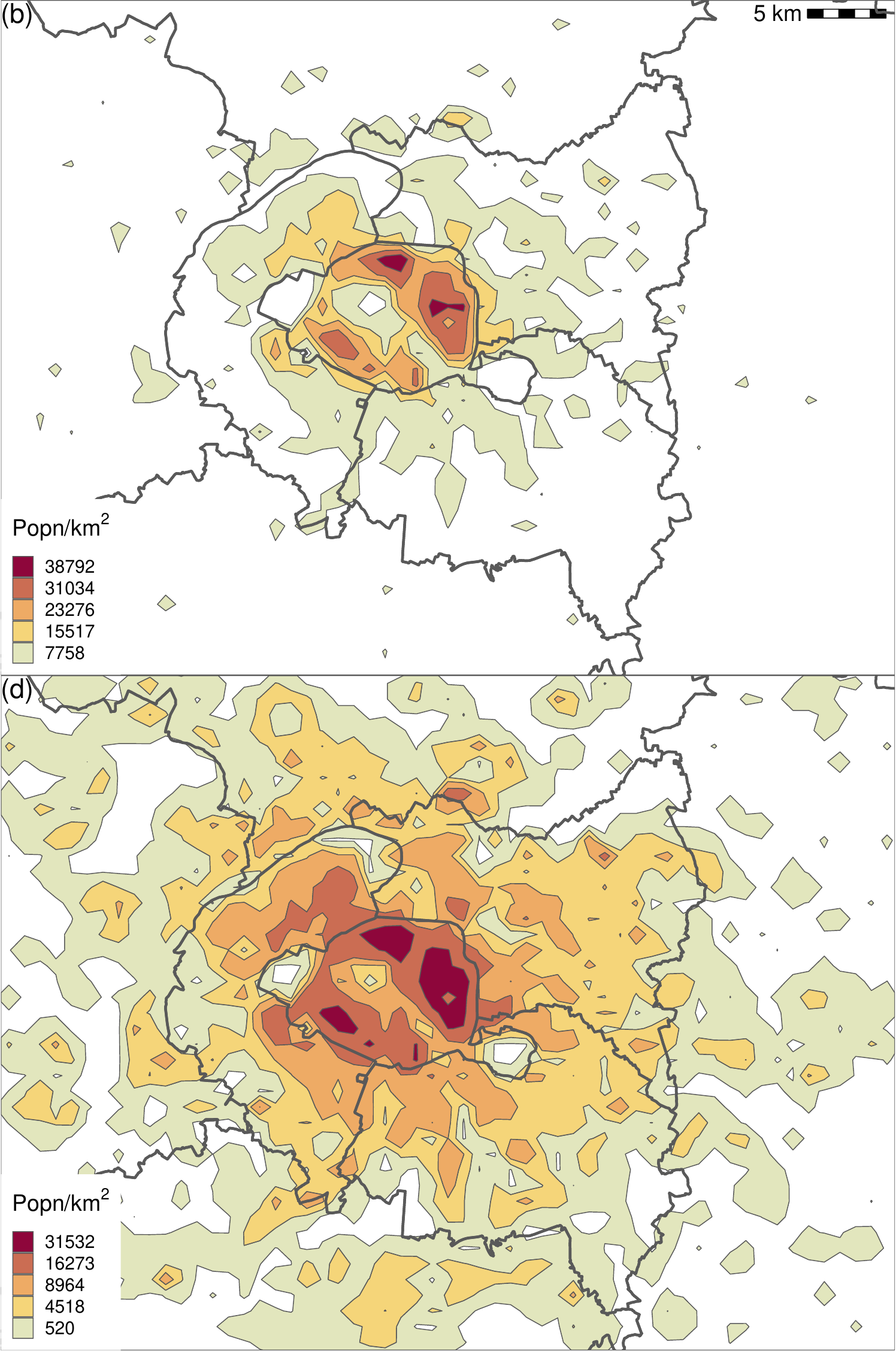}
\caption{Contour regions for population density for Paris capital region on 1~km $\times$ 1~km grid, with  discretised sequential `Heat' colour scale. (a) Equal length contour levels. (b) Natural (Jenks) contour levels.}
\label{fig:idf-popn3}
\end{figure}

The equal length contour regions are 38\,792 (red), 31\,034 (dark orange), 23\,276 (mid orange), 15\,517 (light orange), 7\,758 (yellow) in Figure~\ref{fig:idf-popn3}(a). These contour levels  emphasise the higher population densities: the higher levels delimit well the hotspots, but the lower levels do not delimit the extent of the spatial population distribution as well as the lower density contour levels in Figure~\ref{fig:idf-popn2}(a). 
The natural contour levels 31\,532 (red), 16\,273 (dark orange), 8\,964 (mid orange), 4\,518 (light orange), 520 (yellow) in Figure~\ref{fig:idf-popn2}(d) reveal a similar contour plot with density contour levels, though the lower levels give the visual impression that the population distribution is more evenly spread in the peripheral areas of the ROI. 

\begin{figure}[!ht]
\centering
\includegraphics[width=\textwidth]{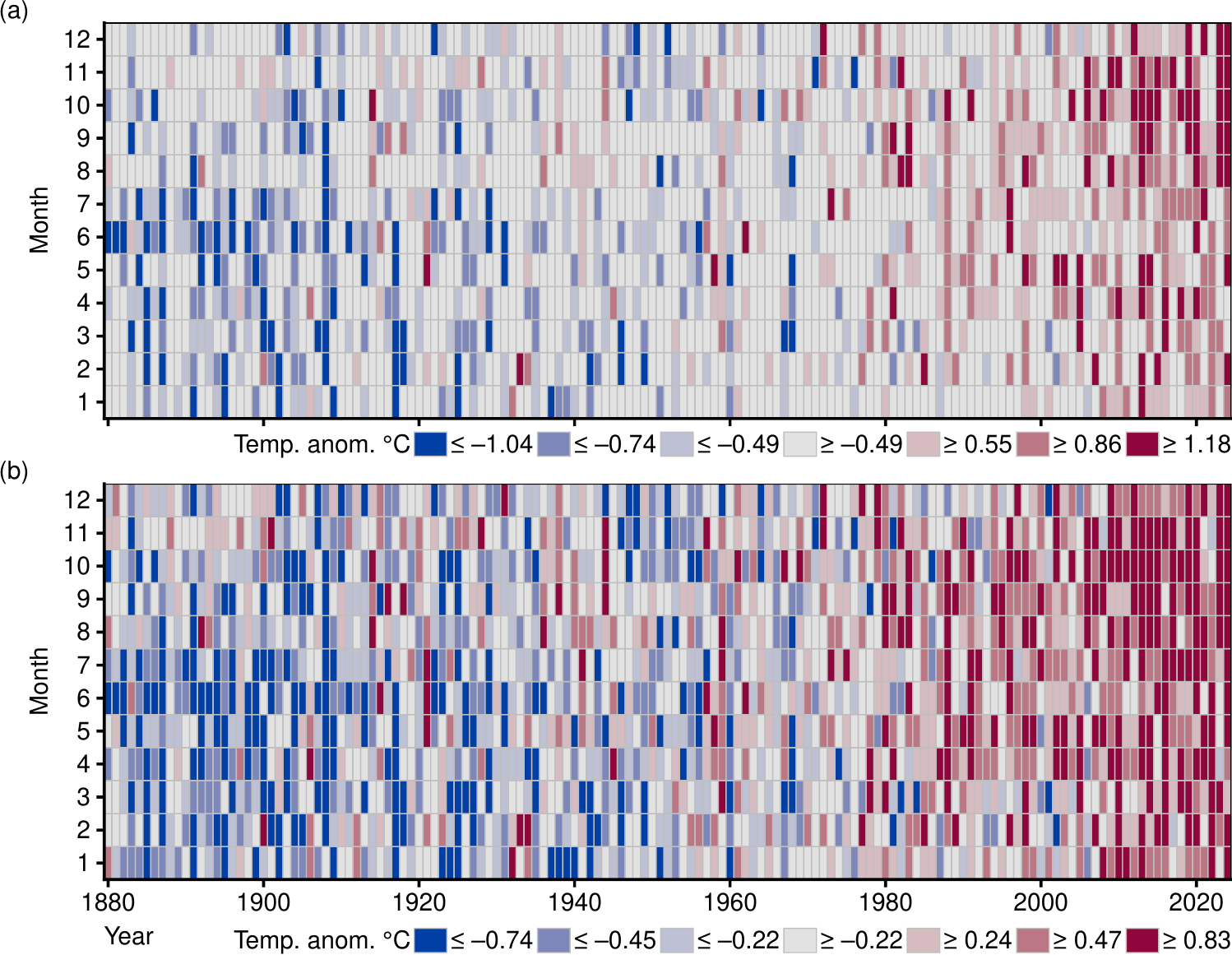}
\caption{Heat maps for year-monthly temperature anomaly time series, with discretised diverging `Red-Blue' colour scale. (a) Density contour levels. (b) Naive quantile contour levels.}
\label{fig:temp-anom3}
\end{figure}

\begin{figure}[!ht]
\centering
\includegraphics[width=\textwidth]{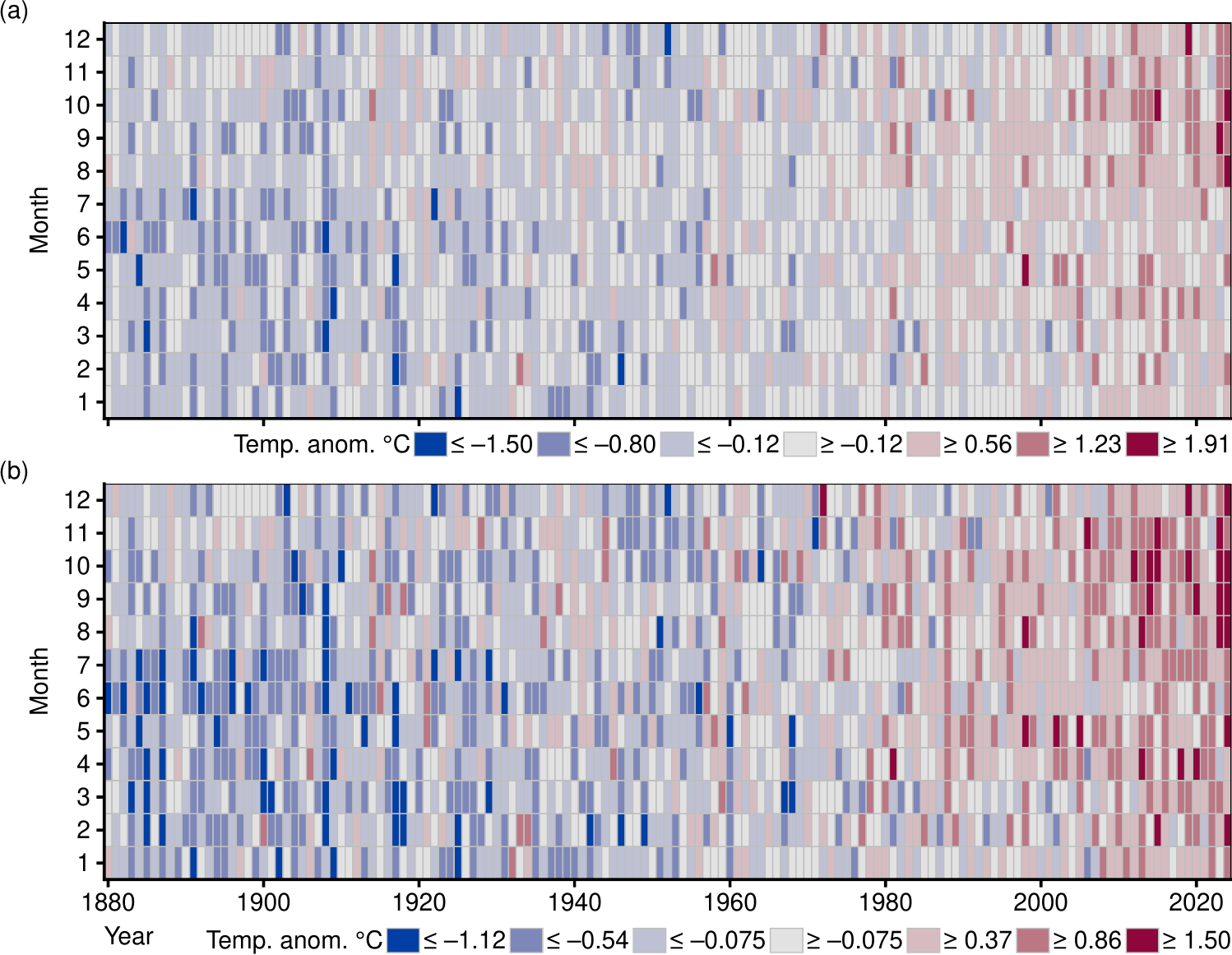}
\caption{Heat maps for year-monthly temperature anomaly time series, with discretised diverging `Red-Blue' colour scale. (a) Equal length contour levels. (b) Natural (Jenks) contour levels.}
\label{fig:temp-anom4}
\end{figure}

For the year-monthly temperature anomaly times series in Western Australia, we are able to compute the analogous contour levels. Since these are non-spatial gridded data, we do not compute contour regions as above and instead we visualise the time series heat maps with colours corresponding to the contour levels. In Figure~\ref{fig:temp-anom3}(a) are the density contour levels, $\hat{g}^-_{0.25}= -1.04, \hat{g}^-_{0.5}=-0.74, \hat{g}^-_{0.75}=-0.49, \hat{g}^+_{0.75}=0.55, \hat{g}^+_{0.5}=0.86, \hat{g}^+_{0.25}=1.18$. There is a large interval of anomalies in $[-0.49^\circ\mathrm{C}, 0.55^\circ\mathrm{C}]$ that are considered to be neutral (grey), so the warming trends (above 1.18$^\circ\mathrm{C}$ in dark red) since the 2000s and in the second half of the year become more prominent. For the naive quartile contour levels in Figure~\ref{fig:temp-anom3}(b), most grid cells are coloured in mid/dark blue or mid/dark red, which provides a less concise summary of the warming trends.
The equal length contour levels in Figure~\ref{fig:temp-anom4}(a) gives the impression of more gradual warming trends since most grid cells are coloured in light blue, grey or light red. The natural contour levels Figure~\ref{fig:temp-anom4}(b) is the most similar to the density contour levels, though the warming trends are less prominent due to the larger number of mid blue/mid red grid cells spread more evenly through the time series heat map.

For these experimental data sets, we do not have the proxies for the ground truth like for the simulated data to which we can compare the estimated contour regions. Nonetheless, to gain insight into the robustness of these estimation methods, we perform a simple sensitivity analysis. Let $\bX_1, \dots, \bX_M$ be the experimental data set with $M$ grid cells, and let the $\tau$-level density levels/contours be $\hat{f}_\tau,\hat{R}_\tau$. These are reasonable targets since from the simulated data, these density contours were the best performing estimates. We generate a simulated replicate in each grid cell from a model of the experimental data: for the Paris population data, this is a Poisson model with mean equal to the observed population, and for the temperature anomaly data, a Gaussian model with mean equal to the observed temperature anomaly and standard deviation equal to 0.25$^\circ\mathrm{C}$. From this data replicate, we compute the density contour, naive quantile, equal length and natural contour levels/regions. For the Paris population data, we compute the error of the symmetric difference as in the simulation study, except that the proxies are replaced by the original density contour regions $\hat{R}_{0.9}, \dots, \hat{R}_{0.1}$. For the temperature anomaly time series, contour regions are not required, so we compute the errors to the original density contour levels $\hat{g}^-_{0.25}, \hat{g}^-_{0.5}, \hat{g}^-_{0.75}, \hat{g}^+_{0.75}, \hat{g}^+_{0.5}, \hat{g}^+_{0.25}$. We carry out 100 replicates, and the errors are summarised in Tables~\ref{tab:err1}--\ref{tab:err2}. 

Since the targets are the density contour levels/regions of the original data set, then we expect that the mean errors of the density contour levels/regions of the replicates (first row) to be close to them. The objective of this sensitivity analysis is gleaned more from the variability of these errors of the contour levels/regions. The naive quantile contours have the lowest variation because their contour levels are highly biased towards zero for almost all cases: so they are robust but with consistently large bias. On the other hand, the equal length contours are the most variable. Next are the natural contours, which are more robust. The density contours have the lowest variation, apart from the highly biased naive quantile contours. So they are robust to small changes in the input gridded data.

\begin{table}[!htp]
\centering\setlength{\tabcolsep}{4pt}
\begin{tabular}{@{}llrrrrrrr@{}}
\toprule
\thead{Estimator} & \thead{$\hat{R}_{0.9}$} & \thead{$\hat{R}_{0.7}$} & \thead{$\hat{R}_{0.5}$} & \thead{$\hat{R}_{0.3}$} & \thead{$\hat{R}_{0.1}$} \\
\midrule
\multicolumn{6}{@{}l}{\textit{Paris population}} \\[0.5ex] 
Density contour & 0.002$\pm$0.001 & 0.004$\pm$0.001 & 0.005$\pm$0.002 & 0.004$\pm$0.002 & 0.000$\pm$0.000 \\ 
Naive quantile  & 0.074$\pm$0.000 & 0.126$\pm$0.000 & 0.030$\pm$0.000 & 0.170$\pm$0.000 & 0.371$\pm$0.000 \\ 
Equal length    & 0.289$\pm$0.003 & 0.089$\pm$0.003 & 0.111$\pm$0.003 & 0.010$\pm$0.002 & 0.004$\pm$0.002 \\ 
Natural         & 0.075$\pm$0.002 & 0.124$\pm$0.002 & 0.033$\pm$0.005 & 0.028$\pm$0.004 & 0.008$\pm$0.002 \\ 
\bottomrule
\end{tabular} 
\caption{Errors of estimates of contour regions for the Paris population data. Each row is contour region estimation method: density contour, naive quantile, equal length, natural. The columns are  original contour regions $\hat{R}_{0.9}, \hat{R}_{0.7}, \hat{R}_{0.5}, \hat{R}_{0.3}, \hat{R}_{0.1}$. Each table entry is the mean $\pm$ SD of error of symmetric difference of estimated and original contour regions.}
\label{tab:err1}
\end{table}

\begin{table}[!htp]
\centering\setlength{\tabcolsep}{4pt}
\begin{tabular}{@{}lrrrrrr@{}}
\toprule
\thead{Estimator} & \thead{$\hat{g}^-_{0.25}$} & \thead{$\hat{g}^-_{0.5}$} & \thead{$\hat{g}^-_{0.75}$} & \thead{$\hat{g}^+_{0.75}$} & \thead{$\hat{g}^+_{0.5}$} & \thead{$\hat{g}^+_{0.25}$}\\ 
\midrule
\multicolumn{6}{@{}l}{\textit{Temperature anomaly}} \\[0.5ex]  
Density contour & 0.089$\pm$0.026 & 0.066$\pm$0.016 & 0.036$\pm$0.014 & 0.035$\pm$0.013 & 0.053$\pm$0.019 & 0.085$\pm$0.026 \\ 
Naive quantile & 0.243$\pm$0.015 & 0.266$\pm$0.013 & 0.260$\pm$0.012 & 0.308$\pm$0.013 & 0.341$\pm$0.014 & 0.310$\pm$0.018 \\
Equal length & 0.509$\pm$0.158 & 0.131$\pm$0.104 & 0.366$\pm$0.129 & 0.110$\pm$0.077 & 0.444$\pm$0.143 & 0.829$\pm$0.164 \\ 
Natural& 0.180$\pm$0.055 & 0.120$\pm$0.054 & 0.341$\pm$0.060 & 0.234$\pm$0.068 & 0.065$\pm$0.049 & 0.333$\pm$0.097 \\ 
\bottomrule
\end{tabular}
\caption{Errors of estimates of contour levels for the temperature anomaly times series. Each row is contour level estimation method: density contour, naive quantile, equal length, natural. The columns are original contour levels $\hat{g}^-_{0.25}, \hat{g}^-_{0.5}, \hat{g}^-_{0.75}, \hat{g}^+_{0.75}, \hat{g}^+_{0.5}, \hat{g}^+_{0.25}$. Each table entry is the mean $\pm$ SD of difference of estimated and original contour levels.}
\label{tab:err2}
\end{table}

\section{Conclusion}

Gridded data are an important data format for the dissemination and storage of open data. Density contour levels are widely deployed to create interpretable visualisations, and we have introduced algorithms to compute density contour levels for any type of numerical gridded data. We demonstrated that for synthetic and experimental gridded data, our proposal leads to more interpretable visualisations than the current contour level selection methods. These algorithms are available in \pkg{R} packages on the official CRAN repository to facilitate their wider deployment for statistical data analysis.

\bibliographystyle{chicago}

\end{document}